\documentclass[preprint,tightenlines,showpacs,preprintnumbers,amsmath,amssymb,prb]{revtex4}

\usepackage{graphicx}
\usepackage{bm}

\begin{document}
\draft
\title{Chiral Symmetry of Double-Walled Carbon Nanotubes detected in First-principles Optical Absorption Spectra}
\author{Xiaoping Yang,$^{1,2}$ Jiangwei Chen,$^{1}$ Hui Jiang$^{1}$
and Jinming Dong $^{1}$}

\affiliation{$^{1}$National Laboratory of Solid State
Microstructures, and Department of Physics, Nanjing University,
Nanjing 210093, P. R. China\\$^{2}$Department of Physics, Huainan
Normal University, Huainan, Anhui Province 232001, P. R. China }

\begin{abstract}
The linear polarizability absorption spectra of the double-walled
carbon nanotubes (DWNTs) have been calculated by using the
tight-binding (TB) model and sum-over-state (SOS) method,
supplemented by the first principles CASTEP calculations. It is
found that the chiral symmetries of both outer and inner tubes in
the DWNTs can always be identified distinctly by the
characteristic peaks in the absorption spectra of the DWNTs, no
matter what kind of the outer tube is, offering a powerful
experimental tool to measure
precisely the chiral angle of the inner tube of a DWNT.\\
\end{abstract}


\date{2 March 2007}
\maketitle

In the past decade carbon nanotubes (CNTs) [1-3], both
single-walled (SWNT) and multi-walled (MWNT) had been extensively
investigated, which is motivated by their special electrical and
mechanical properties as well as by their potential applications
in future's nanostructured materials. Their unique electronic
structures and mechanical properties are proved to be a rich
source of new fundamental physics and also make them promising
candidates as nanoscale quantum wires, single electron and
field-effect transistors and sensors.

The SWNT is composed of a rolled up 2D-graphite sheet, and
discovered first by Iijima's group in 1991. The carbon atoms on a
SWNT are arranged on a helical line around its axis. The
geometrical structure of the SWNT defined by a pair of intergers
($n, m)$ determines its radius and chirality, and so entirely its
electronic structure, and optical property. It is known that the
SWNTs with $n-m$ being a multiple of 3 are metallic, and all
others are semiconducting [4].

The DWNT is the simplest type of MWNT, which has been made
experimentally by many kinds of methods, and consists of two
concentric cylindrical graphene layers. The nucleation of the
inner tube should occur after the growth of the outer tube
according to the 'yarmulke mechanism' [5], which means that the
inner tube diameter should be determined by the outer tube one.

Early studies on DWNTs focused theoretically on their electronic
structure, stability, etc. [6--9]. A DWNT can be composed of a
pair of inner and outer constituent layers with any chiralities,
making different kinds of DWNTs, such as metal--metal,
metal--semiconducting, or semiconducting--semiconducting
nanotubes. It was shown that the band structure of a DWNT depends
on the combined configurations of the inner and outer tubes [7],
but their stability depends only on their interlayer spacing [6].

On the other hand, it is of fundamental and practical importance
to characterize the detailed geometrical structure of a DWNT
because it has a close relationship with the electronic,
mechanical and optical properties of the DWNT. It is in practice,
however, a very difficult task to determine the geometrical
structure of a DWNT, especially its inner tube. Usually, both
scattering tunneling spectroscope (STS) and scattering tunneling
microscope (STM) can be used to roughly measure the outer tube
radius, and so approximate radius of the inner tube. However, the
chiral angle of the inner tube can not be known by this kind of
measurements. The Resonant Raman Scattering and selected area
electron diffraction (SAED) measurements combined with the
theoretical simulations have also been used to determine the
chiral indices of SWNTs, and are in development to get information
about those of the inner tubes in DWNTs [10,11].

It is well known that the optical absorption is a fundamental
property of a solid and can be easily and exactly measured
experimentally. In general, the optical property can be well
described by the electronic energy band structure of the material,
which has a close relation with its geometrical structure,
especially in the nanostructured materials. In this paper, we
investigate how the chiral indices of the inner tubes of DWNTs can
be identified unambiguously by an analysis of the optical spectra
of the DWNTs.

We firstly use a TB model with one $\pi $-orbital per carbon atom
to get the electronic structures of the DWNTs, which has been
successfully applied to describe the SWNT, DWNT, and MWNT [12-14],
and then use the SOS approach to calculate the linear absorption
spectra [15,16] of DWNTs with finite lengths in real space.

The TB Hamiltonian is given by [12-14]
\begin{equation}
\label{eq1} H = \gamma _0 \sum\limits_{i,j} c_j^{\dag } c_i -
W\sum\limits_{{i'},{j'}}\cos(\theta _{{i'}{j'}})e^{(a
-d_{i'j'})/\delta}c_{j'}^{\dag} c_{i'},
\end{equation}
  where $\gamma _0 $ ($= -2.75$ eV) is the intra-layer hopping
parameter between the nearest neighbor sites, $i$ and $j$, and $W$
is the inter-wall interactions strength between interlayer sites,
$i'$ and $j'$, with a distance of $d_{i'j'} $ and the cutoff of
$d_{i'j'}$ $>$ 3.9 {\AA}. Here $\theta _{i'j'}$ is the angle
between two $\pi $-orbitals on the sites $i'$ and $j'$, and
$c_{i}$ is the annihilation operator of an electron on site $i$.
$a$ (= 3.34 {\AA}) is the distance between two carbon walls, and
$\delta $ = 0.45 {\AA}. As well known, the parameters used in the
TB model are usually obtained by fitting the TB calculation
results with those of the ab initio calculations, or the
experimental data. Estimation based on ab initio calculations of
$W$ gives $W$ (= $\gamma _0/8)$. The weak inter-layer interaction
is caused by the larger inter-layer distance of 3.4 {\AA},
compared with intra-layer bond length of 1.42 {\AA}.

Within the independent electron approximation and SOS approach,
the linear polarizability $\alpha (\omega )$ is expressed as [15]
\begin{equation}
\label{eq2}
\alpha (\omega ) = 2\sum\limits_{\begin{array}{l}
 n \in occ \\
 p \in unocc \\
 \end{array}}{\mu _{np} \mu _{pn} (\frac{1}{\varepsilon _{pn} - \omega } +
\frac{1}{\varepsilon _{pn} + \omega })}
\end{equation}
In the above formula, $\varepsilon _{np} = \varepsilon _n -
\varepsilon _p $, and $\mu _{nm} $ is the dipole transition matrix
elements between the one-electron state $Z_{n,s} $ and $Z_{m,s} $,
which is given by
\begin{equation}
\label{eq3}
\left\langle n \right|\mu _\alpha \left| m \right\rangle = \sum\limits_{j,s}
{Z_{n,s}^\ast (j)( - er_j )} Z_{m,s} (j).
\end{equation}
Here, $r_j$ is the coordinate of the $j_{th}$ atom. The linear
polarizability absorption spectra is the imaginary part of the
linear polarizability $\alpha (\omega )$.

Then, we have also made the first-principle calculations on the
polarized absorption spectra of the DWNT by the CASTEP code [17],
which is compared with the results obtained by the TB method.

The nanotubes are usually treated as infinite periodic one
dimensional crystals in the theoretical studies. But in practical
experiments, the nanotubes have finite lengths. So, it is
interesting to study physical properties of the carbon nanotubes
with finite lengths, which has big effects on their electronic
structures and corresponding optical responses [16,18]. For
example, the finite length ($n, m)$ SWNTs with $n-m$ being not
multiples of 3 will have smaller linear polarizability $\alpha
(\omega )$ [16].

In our calculations, the nanotube is cut perpendicular to its
axis, making it to have two open ends. Taking into account the
computational time and cost, we select the calculated CNT length
of up to 172.2 {\AA}, which is equivalent to that of 70 periods of
the armchair tube (12, 12). Correspondingly, the calculated energy
region for the optical spectrum is restricted in the 0.4 $\sim $ 4
eV. The interlayer distance of DWNTs is usually taken as 3.4
{\AA}.

Firstly, we choose two kinds of SWNT as the possible outer tube of
the DWNTs: armchair tube (12, 12) (radius, $r$ = 8.136 {\AA}) and
chiral semiconductor tube (13, 11) ($r$ = 8.134 {\AA}). Their
absorption spectra $\alpha _\parallel \left( \omega \right)$
polarized to tube direction are calculated in the TB approximation
and shown in Figs. 1(a), 1(b), respectively. And their
corresponding densities of states (DOSs) are shown in Figs. 1(c)
and 1(d), respectively. It is seen that the characteristic
absorption peaks labeled by symblo '$o$' lie at 1.45, 2.8 and 3.9
eV for (12, 12) tube, and at 0.5, 1.0, 1.9, 2.3, 3.2 and 3.5 eV
for (13, 11) tube, which are contributed by the 1$^{st}$,
2$^{nd}$, 3$^{rd}$, etc, pair of Van Hove Singularity (VHS) DOS
peaks. For example, it can be clearly seen that the distances
between the 1$^{st}$ pair of VHS DOS peaks for (12, 12), and (13,
11) tubes are about 1.45 eV, and 0.5 eV, respectively, which
correspond exactly to the positions of two 1$^{st}$ absorption
peaks for both SWNTs [seen from Figs. 1(a) and 1(b)]. The
characteristic absorption peaks will become more evident when the
calculated CNT length is taken to be longer.

For these two outer tubes, different possible inner tubes are
chosen and studied as follows.

The inner tube is armchair (7, 7) tube ($r$ = 4.746 {\AA}). In
this case, two absorption spectra $\alpha _\parallel \left( \omega
\right)$ for two DWNTs, commensurate (7, 7)@(12, 12), and
incommensurate (7, 7)@(13, 11), are calculated in the TB
approximation and shown in Fig. 2 , in which are also shown the
corresponding results for pure (7, 7) tube. In order to define
more directly the characteristic absorption peaks of the inner
tube, we also show the joint density of states (JDOS) in Fig. 2(a)
for the vertical transition between the same band index. The only
one sharp peak at 2.4 eV in the JDOS of the inner (7, 7) tube
comes from the 1$^{st}$ pair of the VHS peaks in its DOS, which
induces one sharp absorption peak at 2.4 eV, shown in Fig. 2(b)
and labeled by a symbol '$i$'. It is seen by comparing Fig. 1 and
Fig. 2 that all the characteristic absorption peaks of both outer
and inner tubes can be identified distinctly in the absorption
spectra [Figs. 2(c) and 2(d)] of the DWNTs with an allowable shift
of about $\pm 0.1$ eV, which is caused by the splitting of the
degenerated bands, induced by the weak interlayer coupling in the
DWNT. For example, the 1$^{st}$ absorption peak of the inner (7,
7) tube can be found clearly in the absorption spectra of both
DWNTs [Figs. 2(c) and 2(d)] although it happens to overlap with a
peak of the outer tube (13, 11) at 2.4 eV in the DWNT (7, 7)@(13,
11).

We can also choose a chiral metal tube (11, 2) ($r$ = 4.746 {\AA})
as the inner tube, whose period is equal to seven times of that
for (12, 12) tube, and incommensurate to that of tube (13, 11).
The calculated JDOS of SWNT (11, 2) is shown in Fig. 3(a), and the
absorption spectra $\alpha _\parallel \left( \omega \right)$ for
pure tube (11, 2), the commensurate DWNT (11, 2)@(12, 12), and
incommensurate DWNT (11, 2)@(13, 11) are given in Figs. 3(b), 3(c)
and 3(d), respectively. Since each VHS peak of the inner chiral
metal tube (11, 2) is now split into two with much different
heights, there appear two sharp JDOS peaks in Fig. 3(a), which
induce two big absorption peaks at 2.3 and 2.6 eV, shown in Fig.
3(b). Both of them can also be identified distinctly in the
absorption spectra of the two DWNTs [Figs. 3(c) and 3(d)] with the
same shift of about $\pm 0.1$ eV. In the same time, the
characteristic absorption peaks of outer tubes (12, 12) and (13,
11) can be found clearly in the absorption spectra of two DWNTs.
So, it is obvious that the chiral indices of a chiral or achiral
inner metal tube can be identified well by the absorption spectra
of corresponding DWNTs no matter the outer tube is metal or
semiconductor.

Above, both metal tubes, one (7, 7) and another (11, 2) are chosen
as the inner tube of the DWNTs. Now, we choose a chiral
semiconducting tube (8, 6) ($r$ = 4.762 {\AA}) as the inner tube.
Its JDOS is given in Fig. 4(a). The calculated absorption spectra
$\alpha _\parallel \left( \omega \right)$ for pure (8, 6) tube,
the incommensurate DWNTs (8, 6)@(12, 12) and (8, 6)@(13, 11) are
shown in Figs. 4(b), 4(c), and 4(d), respectively. Now, the more
JDOS peaks of the inner (8, 6) tube appear in the energy window of
0 to 4 eV, producing thus more characteristic absorption peaks at
0.85, 1.65, 3.2 and 3.7 eV in Fig. 4(b). Even so, it is clearly
seen from Figs. 4(c) and 4(d) that these four absorption peaks of
the inner tube (8, 6) can always be well identified although again
the 3$^{rd}$ peak coincides with one absorption peak of outer tube
(13, 11) in Fig. 4(d).

From the above results, it is known that no matter the inner tube
is metal or semiconductor, achiral or chiral, its characteristic
absorption peaks can always be identified clearly from the
absorption spectra of the corresponding DWNTs, which is
independent of selections of the outer tubes, indicating the
inter-tube coupling has only minor effects on the absorption
spectra of a DWNT. So, we can conclude that it is possible
experimentally to deduce the chiral indices of the inner tube for
any DWNTs by comparing their absorption spectra with those
different types of SWNTs.

Finally, in order to check correctness of our TB calculations, we
have also made the first-principle calculations in the framework
of local density approximation (LDA) on the polarized absorption
spectra $\alpha _\parallel \left( \omega \right)$ of the DWNT (7,
7)@(12, 12) and related SWNTs (7, 7) and (12, 12) by the CASTEP
code [17], in which the exchange-correlation energy of the
Ceperley and Alder form [19] was used. The ion-electron
interaction is modeled by ultra-soft local pseudopotentials of the
Vanderbilt form [20] for the carbon atoms with maximum plane wave
cut-off energy of 240 eV. In the calculation, a supercell geometry
[21] is used in which the tubes are aligned in a hexagonal array
with an adjacent inter-tube distance of 28 {\AA}, being larger
enough to prevent the tube-tube interactions. The space group is
$P_1 $ for the computational models. The Monkhorst-Pack scheme
[22] with a distance of 0.03/{\AA} between points is used for the
sampling in reciprocal space. The calculated results are given in
Fig. 5. The absorption peaks of SWNT (7, 7) are found to be at
2.45, 3.44 and 3.84 eV, and those for SWNT (12, 12) are at 1.44,
2.09, 2.54, 2.79, 3.19 and 3.94 eV. And those peaks for DWNT (7,
7)@(12, 12) lie at 1.38, 1.87, 2.04, 2.45, 2.74, 3.10, 3.37, 3.53,
3.78 and 4.03 eV. By comparison, we find that the absorption peaks
of both the outer and inner tube can be all identified distinctly
in the absorption spectra of the DWNT with an allowable shift of
about $\pm 0.1$eV, which comes also from the interlayer
interaction of the DWNT.

In addition, comparing Figs. 5(a) and 5(b) with Fig. 1(a) and Fig.
2(b), we know that the characteristic absorption peaks of pure (7,
7) and (12, 12) tubes obtained by the TB model and SOS method can
all be found in their absorption spectra got from the
first-principles calculations, which lie at 2.45 eV for (7, 7)
tube, and at 1.44, 2.79, and 3.94 eV for (12, 12) tube with an
allowable shift of about $\pm 0.05\mbox{ eV}$. So, we conclude
that the obtained results by the TB model and SOS method are much
closer to those by the first-principles method, especially for the
absorption peak positions, demonstrating that both of the TB model
and SOS method are suitable for the calculations of the linear
optical absorptions for the DWNTs and SWNTs.

In summary, it is shown by the numerical calculations that the
optical responses can be used to identify clearly the chiral
symmetries of the outer and inner tubes in the DWNTs, based upon
which a powerful experimental tool can be proposed for measuring
unambiguously the chiral angle of the inner tube of a DWNT.\\

This work was supported by the Natural Science Foundation of China
under Grant No.10074026, No. 90103038 and No. A040108. The authors
acknowledge also support from a Grant for State Key Program of
China through Grant No.1998061407.

\newpage

\begin{figure}[htbp]
\includegraphics[width=0.8\columnwidth]{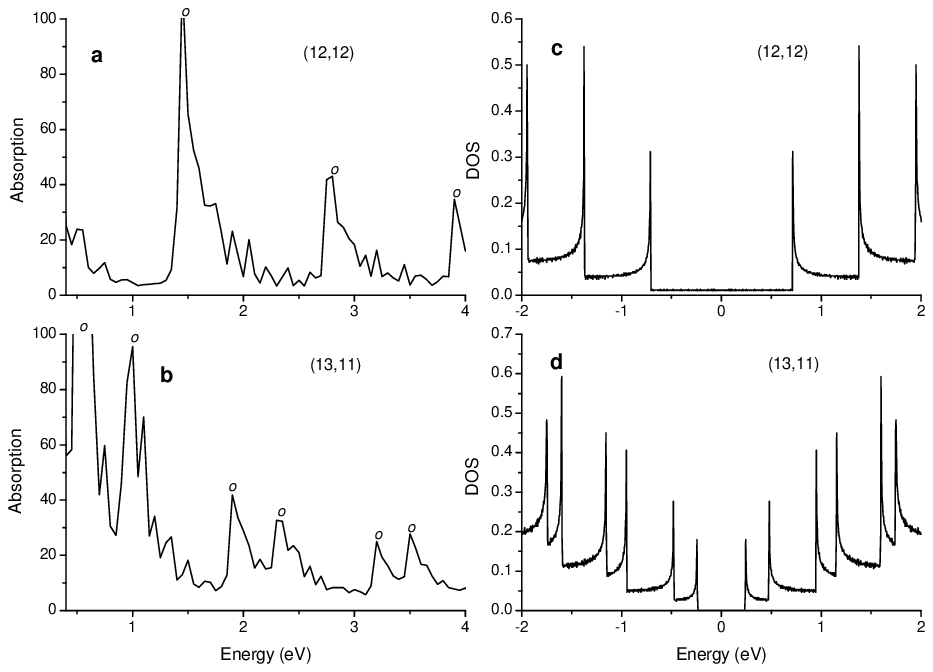}
\label{fig1} \caption{ The absorption spectra $\alpha _\parallel
\left( \omega \right)$ (in {\AA}$^{3}$ ) for (a) (12, 12) tube , (b)
(13, 11) tube and corresponding density of states for (c) (12, 12)
tube, (d) (13,11) tube. The characteristic absorption peaks of the
outer and inner tubes are labeled by the symbol '$o$' and '$i$',
those are the same in the other figures of this paper.}
\end{figure}

\begin{figure}[htbp]
\includegraphics[width=0.8\columnwidth]{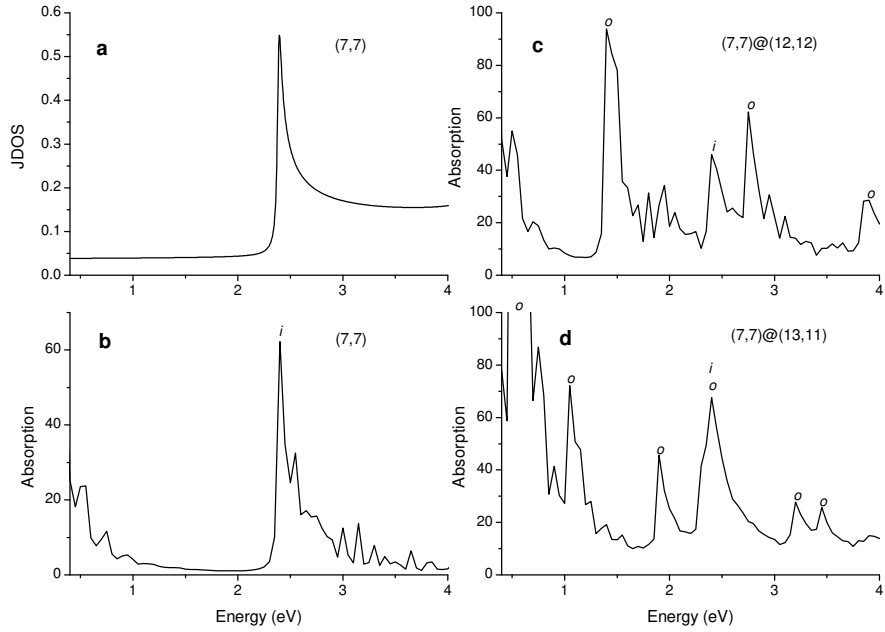}
\label{fig2} \caption{(a) JDOS of (7, 7) tube. The absorption
spectra $\alpha _\parallel \left( \omega \right)$ (in {\AA}$^{3}$ )
for (b) pure (7, 7) tube, (c) DWNT (7, 7)@(12, 12), and (d) DWNT (7,
7)@(13, 11).}
\end{figure}

\begin{figure}[htbp]
\includegraphics[width=0.8\columnwidth]{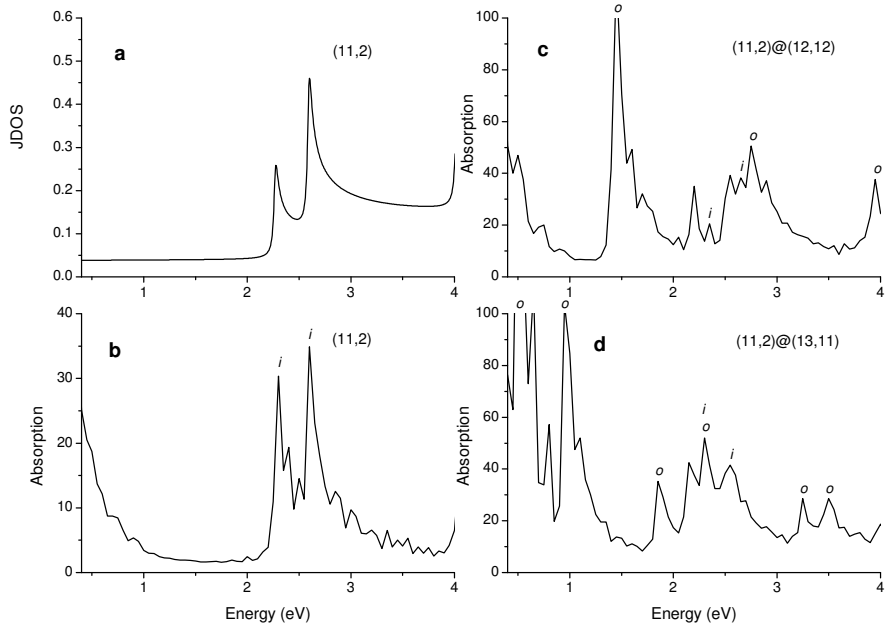}
\label{fig3} \caption{(a) JDOS of pure (11, 2) tube. The absorption
spectra $\alpha _\parallel \left( \omega \right)$ (in {\AA}$^{3}$ )
for (b) pure (11, 2) tube, (c) DWNT (11, 2)@(12, 12), (d) DWNT (11,
2)@(13, 11).}
\end{figure}

\begin{figure}[htbp]
\includegraphics[width=0.8\columnwidth]{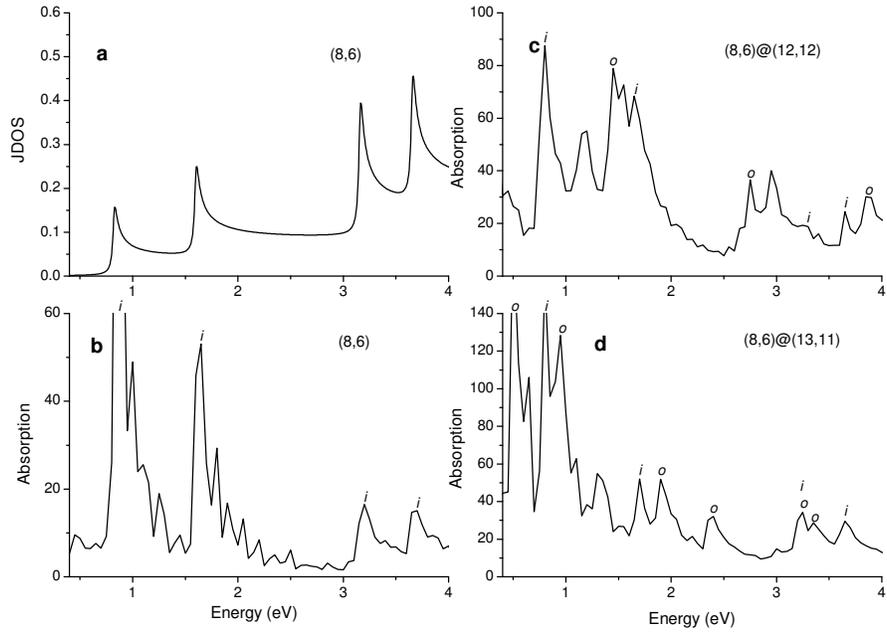}
\label{fig4} \caption{(a) JDOS of pure (8, 6) tube. The absorption
spectra $\alpha _\parallel \left( \omega \right)$ (in {\AA}$^{3}$ )
(b) pure (8, 6) tube, (c) DWNT (8, 6)@(12, 12), and (d) DWNT (8,
6)@(13, 11).}
\end{figure}

\begin{figure}[htbp]
\includegraphics[width=0.8\columnwidth]{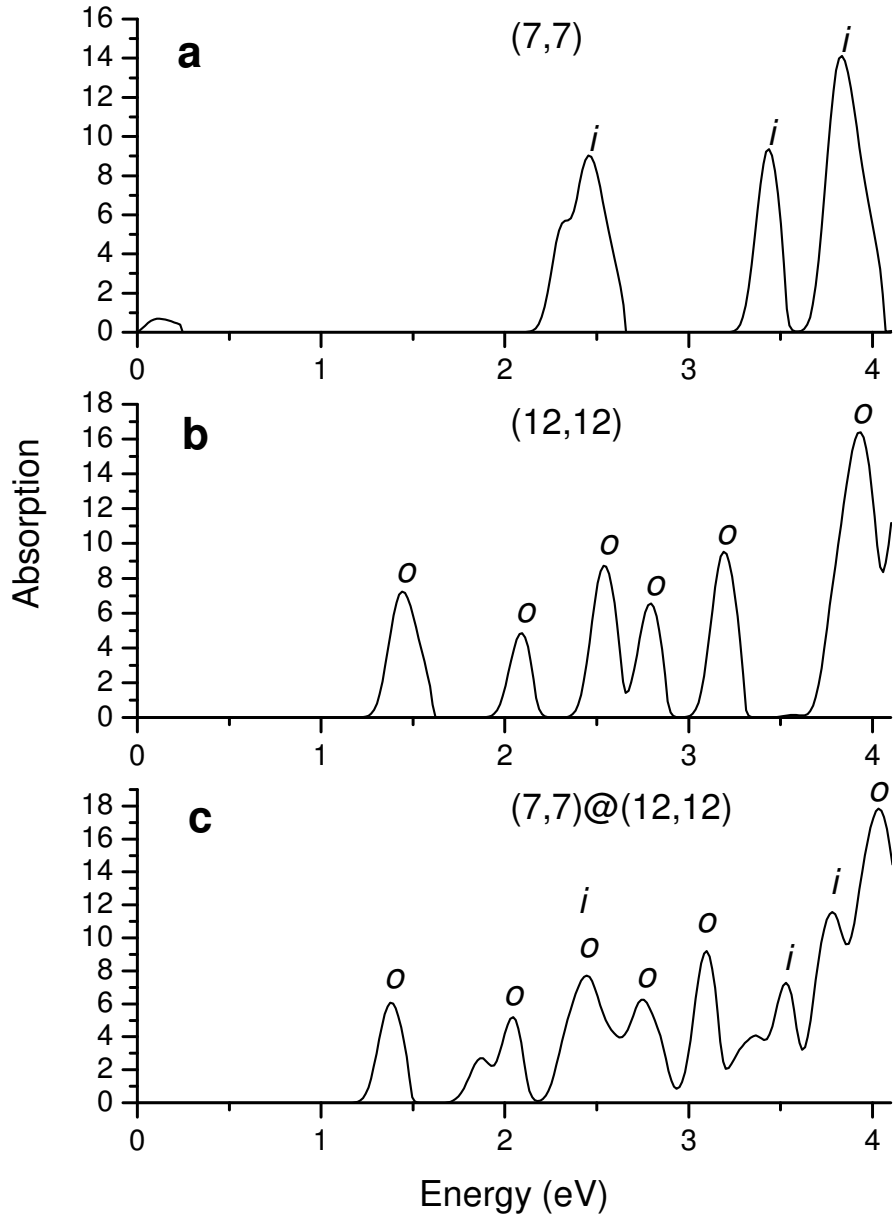}
\label{fig5} \caption{The absorption spectra $\alpha _\parallel
\left( \omega \right)$ (in {\AA}$^{3})$ for (a) pure (7, 7) tube,
(b) pure (12, 12) tube, and (c) DWNT (7, 7)@(12, 12) obtained by the
first-principles calculation.}
\end{figure}

\end{document}